\documentclass[]{spie}  

\usepackage{amsmath,amsfonts,amssymb}
\usepackage{graphicx}

\title{Speckle Imaging Through a Coherent Fiber Bundle}

\author[a]{James W. Davidson Jr.}
\author[b]{Elliott P. Horch}
\author[a]{Steven R. Majewski}
\affil[a]{Department of Astronomy, University of Virginia, P.O. Box 400325, Charlottesville, VA USA 22904-4325}
\affil[b]{Department of Physics, Southern Connecticut State University, 501 Crescent St., New Haven, CT USA 06515-1330}

\authorinfo{Further author information: (Send correspondence to J.W.D.Jr.)\\J.W.D.Jr.: E-mail: jimmy@virginia.edu, Telephone: 1 434 924 4909\\  E.P.H.: E-mail: horche2@southernct.edu\\  S.R.M.: E-mail: srm4n@virginia.edu}

\begin{document} 
\maketitle

\begin{abstract}
Speckle imaging is a well known method to achieve diffraction-limited (DL) imaging from ground-based telescopes. The traditional observing method for speckle has been to observe a single, unresolved, source per telescope pointing over a relatively small field of view (FOV).  The need for large DL surveys of targets with high sky density motivates a desire for simultaneous speckle imaging over large FOVs, however it is currently impractical 
to attain this by covering the entire focal plane with fast readout detectors.  
An alternative approach is to connect a relatively small number of detector pixels to multiple interesting targets spanning a large FOV through the use of optical fibers, a technique employed in spectroscopy for decades. However, for imaging we require the use of coherent fiber bundles (CFBs).
We discuss various design considerations for coherent fiber speckle imaging with an eye toward a multiplexed system using numerous configurable CFBs, and we test its viability with a prototype instrument that uses a single CFB to transport speckle images from the telescope focal plane to a traditionally designed, fast readout speckle camera. Using this device on University of Virginia's Fan Mountain Observatory 40-inch telescope
we have for the first time successfully demonstrated 
speckle imaging through a CFB, using
both optical and NIR detectors. Results are presented of DL speckle imaging of well-known close (including subarcsecond) binary stars 
resolved with this fiber-fed speckle system and compared to
both literature results and traditional speckle imaging taken with the same camera directly, with no intervening CFB.
\end{abstract}


\section{INTRODUCTION}
The advent of electron-multiplying CCD (EMCCD) cameras has rejuvenated 
the technique of speckle imaging so that it is now routinely used in
follow-up observations for exoplanet host stars and other large
survey projects (see e.g.\ [\citenum{horch_spie}] and references therein). In addition to the first speckle cameras that used EMCCDs
[\citenum{tok08,horch11}], new instruments based on the same 
technology have recently been completed and are in use at the
WIYN [\citenum{scott18}] and Gemini-N  [\citenum{scott16}] telescopes. 
These newer instruments use EMCCDs with a pixel format of 1024 $\times$ 1024 pixels, much larger than is generally needed for speckle imaging. At the same time, improvements in infrared detectors have improved 
in terms of readout speed and read noise, and are available in 
comparable formats. Often, when such systems are 
used for speckle imaging observations, 
only a comparatively small sub-array is read out and stored
because individual targets subtend small angular dimensions.

On the other hand, there is a growing need for largescale, systematic 
diffraction-limited surveys of targets of sufficiently high sky density that developing techniques for 
simultaneous speckle imaging over large field of views (FOVs) is desirable.
For example, understanding the wide variety of star and planet formation architectures being found by both transit and radial velocity surveys requires high resolution
imaging of large numbers of systems in all evolutionary phases across
Galactic environments.
Diffraction-limited imaging can identify false positive exoplanet
detections, as well as hierarchical star/planet systems, thereby
providing a vast, and well-vetted database to address questions such
as: the frequency and stability of planet, stellar binary, and
multi-star systems; the influence of long period companions on
exoplanet host system architectures; and dependencies of these
architectures on local environmental conditions. 
However, while the {\it Kepler} and K2 missions
have identified thousands of potential companions to nearby stars spanning the mass range from exo-Earths to equal-mass binaries, 
over the next decade the
TESS and {\it Plato} missions are poised to deliver of order $10^5$  detections of exoplanet mass companions alone.
Clearly, the sheer number of interesting follow-up targets being produced by these surveys makes the task of wholesale high resolution imaging characterization much less practical with single-object instruments like the Differential Speckle Survey Instrument (DSSI) [\citenum{Horch1}], NN-EXPLORE Exoplanet Stellar Speckle Imager (NESSI) [\citenum{scott16}], and Robo-AO [\citenum{Ziegler16}].
Huge increases in throughput are needed.
However, it is impractical and costly to attain this by paving telescope focal planes with fast readout detectors.

These developments have led two of us (S.R.M. and J.W.D.) to 
consider alternative methods for multiplexed  
speckle observations of numerous sources observed simultaneously by large field-of-view (FOV) telescopes.
This could be done if, as with
fiber-fed multi-object spectrographs, fiber-optic technology were
used to place the speckle images from different stars across 
the telescope FOV onto different subregions of one (or a small 
number) of the new, large format, fast readout detectors.
The key difference in the speckle imaging
application would be that a coherent fiber bundle (CFB) would be needed
to preserve and transfer the image information from the telescope 
focal plane to the detector active area. If a number of CFBs were
focused onto the detector active area, the pixels 
could all be readout and stored simultaneously. However, the introduction of 
the CFBs could also affect the quality of the speckle 
data, depending on their transmission characteristics.

In this paper, we explore aspects of this proposed next generation
speckle imaging concept. We first lay out various considerations
for introducing CFBs into speckle imaging.
Section 3 then describes a prototype system designed to test some 
of these considerations with a capability to record speckle patterns 
either directly from a telescope or through a CFB. 
Using this device, we demonstrate that speckle imaging is possible 
through a CFB, an important step toward the development of a 
multi-object speckle imager.  The engineering data described were taken 
with a 40-inch telesocpe using both a near-infrared fast-readout camera 
and an EMCCD. These results inform not only design considerations 
for 
a multi-object speckle imaging system but the techniques needed to 
process such data. 
We conclude with some basic comments on the proposed design 
of a fiber-fed speckle system for potential use on
the WIYN 3.5-m telescope.

\section{CONSIDERATIONS FOR COHERENT FIBERS IN SPECKLE IMAGING}

To our knowledge, use of CFBs in speckle imaging has not
been previously explored.  Obviously, the proposition of multiplexed 
speckle imaging using CFBs requires contemplation of 
a number of practical considerations and potential design trade-offs 
on the telescope side. For the present discussion, we only briefly 
touch on these aspects (e.g., \S2.1).  Two  
obvious relevant requirements are that (1) an appropriate 
telescope is available --- i.e., one with large enough aperture to 
be in a compelling diffraction-limited resolution regime, but with large enough FOV 
relative to the sky density of interesting sources to deliver 
sufficient targets per pointing to realize adequate multiplexing 
gains (we argue below that 1-2 orders of magnitude is a compelling
increase in observing efficiency), and (2) that a means of positioning CFBs across the telescope focal plane is feasible.  We note
that numerous modes of positioning single fibers for multifiber
spectroscopy have been developed over the past few decades, and no 
doubt such systems can be made to work with CFBs; 
the MaNGA experiment in Sloan Digital Sky Survey IV [\citenum{Bundy15}]
is one example of a survey making use of multiplexed CFBs.

Our focus here is primarily on the feasibility of introducing coherent
fibers into speckle imaging --- including considerations for multiplexing 
them on the detector end (\S 2.1), and their impact on the detection process 
itself (\S2.2).  
Primary concerns regarding the latter include:
(a) imprint of CFB fiber cellular structure on the speckle Fourier power spectrum,
(b) contrast losses, and (c) light losses.  We have begun testing the concept with simulations (\S2.2) and actual telescopic observations (\S 3-5).  Data collected to date have established the viability of the concept, despite exploring ``worst-case scenarios'' in some regards.

\subsection{Multiplexing Factor Versus Fiber FOV} 

A practical example is useful for setting context and objectives for 
fiber multiplexing. For this exercise we adopt the WIYN 3.5-m
telescope as a reference, given its 1 degree diameter FOV, its tradition in speckle imaging with the Rochester Institute of Technology-Yale Tip-Tilt Speckle Imager (RYTSI), DSSI and NESSI systems [\citenum{Meyer06,Horch1,scott16}], as well as its history of multifiber spectroscopy 
with the Hydra fiber positioner [\citenum{Barden94}].  We explore
its potential use for a comprehensive, diffraction-limited imaging
follow-up survey of targets for NASA's
Transiting Exoplanet Survey Satellite (TESS).

The 450 deg$^2$ TESS Continuous Viewing Zones (CVZs) --- 
the best explored and most valuable parts of the TESS survey --- have
a $\sim$96 deg$^{-2}$ density of candidate target list (CTL) stars, 
with $\sim$53 deg$^{-2}$ classified as ``top priority" [\citenum{Stassun17}].
A telescope like WIYN, with a 1 degree diameter FOV (i.e., covering 0.79 deg$^2$),
could thus deliver data from $\sim$75 CTL stars at a time, 
$\sim$42 of them being high priority.  Efficiently imaging these 
sources could be achieved by packing, say, an 8$\times$8
grid of coherent bundles onto the current 1024$^2$ pixel format
high speed detectors.  Accounting for some losses due to fiber bundle
jacketing, it is not unreasonable to expect that the transmissive
parts of each CFB could mate to an active detector area 
some 100 pixels across.
Imaging at $\lambda$=0.55$\mu$m, the diffraction limit of a
3.5-m telescope is $0.04^{\prime\prime}$.  Therefore, to Nyquist sample the
corresponding speckles would require a CFB core diameter subtending $\le$$0.02^{\prime\prime}$, 
yielding a coherent fiber net FOV of $\lesssim$$2^{\prime\prime}$.  This is 
adequate for detecting companions with separations from the diffraction limit
up to scales where they are obvious in existing or upcoming all-sky imaging surveys, like SDSS or LSST.
Thus, current 
detector array sizes are sufficiently large to accommodate desirable
multiplexing factors, although the pace of
detector development suggests that even larger format arrays, enabling larger CFB FOVs, may be available in the near future.

In terms of target throughput, it is worth mentioning that the
existing (though now several decades old) Hydra positioner is able to
reposition 64 fibers in about 10 minutes.  It can be shown that for
the magnitudes of the TESS CTL sources, 10 minutes of speckle imaging
can attain sufficient $S/N$ to detect stellar companions with 
the contrast currently obtained with the NESSI system.  Thus, 
as a strawman observing system, a 3.5-m telescope with a 
Hydra-like positioner having 64 CFBs could obtain diffraction-limited image
on as many as 192 TESS exoplanet host candidates per hour, 
$\sim$2,000 sources per night, and all of the CTL sources in an
entire CVZ in something like 30 clear, summer nights (the season
when each CVZ is in the sky in its respective hemisphere).
In comparison, the DSSI system (used variously on the 
WIYN, Gemini-N, and Discovery Channel Telescopes) observes 
a total of 50-100 science targets per night depending 
on their brightness. 
Meanwhile, Robo-AO, largely due to its vastly decreased overheads compared to typical AO laser guide star systems, is more efficient, 
but still only capable of observing at best 
20 stars/hour [\citenum{Ziegler17}], or 200 targets per 
summer night.  Thus, the hypothetical multiplexed CFB system
would represent at least an order of magnitude throughput gain
over single source, high resolution imagers.

\subsection{Impact of Coherent Fiber Cell Structure} 

The introduction of a CFB in the speckle imaging lightpath imposes
an additional level of pixelization to that from the 
detector pixels due to the cellular structure given by the network of 
fiber cores. 
Through choice of either the actual physical fiber core 
dimensions of the adopted CFBs, or
appropriate fore- and post-optics, the scale of that pixelization 
can be adjusted.  Ideally, lossless image transfer requires that the 
fiber cores Nyquist sample the speckles on the input side of the CFBs and transfer that sampling to the detector through
an appropriate match of fiber cores to detector pixels. 
In practice, the situation is complicated both by the fact that cellular network of coherent fiber cores is irregularly patterned and the fact that
commercially made CFBs have a limited range of diameters, total fiber core counts and active area fill factors.        

\begin{figure}[ht]
    	\centering
		\includegraphics[scale=1.5]{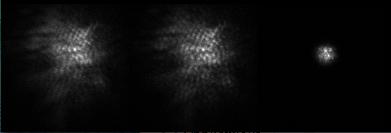}
    	\caption{A 40 ms exposure, $\lambda$=0.88$\;\mu$m speckle pattern from the DCT 4.3m+DSSI. ({\it Left}) Unmasked speckle image. ({\it Center}) The same, but masked with a CFB template, scaled with CFFB core sizes comparable to typical speckle size. Faint speckles  show ``waffling'' due to the mask. ({\it Right})  Rescaled to simulate  $H$-band and with the CFB fiber core sizes reduced to half the speckle width. The image scale is altered to maintain the relative speckle size in pixels, slightly more than Nyquist  sampled, in a 128$^2$ pixel format.   {\it Left} and {\it center} have FOV=$2.3^{\prime\prime} \times 2.3^{\prime\prime}$;  {\it right} has FOV=$6.4^{\prime\prime} \times 6.4^{\prime\prime}$.} 
    	\label{fig:speckles}
\end{figure}

We simulated use of CFBs to transmit speckle images with data from
the DSSI instrument on the 4.3m Discovery Channel Telescope (DCT). These $\lambda$=0.88$\mu$m images were masked with 
scaled templates made from an image of the end of a flatly illuminated coherent bundle. Figure \ref{fig:speckles} shows an 
unmasked speckle pattern ({\it left}) from one 40ms exposure out of a 1000-frame stack, and a masked counterpart ({\it center}), where the mask was scaled to where the fiber cores are comparably-sized to the speckle width.
In the center image, the fainter speckles in particular show a ``waffle'' effect arising from the mask. The same mask was applied to all 1000 frames in the original sequence, and both the unmasked and masked series of images were analyzed similarly.

\begin{figure}[t]        
    	\centering
		\includegraphics[scale=1.5]{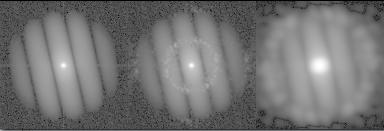}
		\caption{Spatial frequency power spectra for the images in Fig. \ref{fig:speckles}
        The fringe pattern is caused by the presence of the binary companion and  persists in the simulated CFB-transmitted data sets ({\it center, right}). When the fiber cores sizes are reduced in size to Nyquist sample the speckles, the ring of ``noise power'' is moved beyond the diffraction limit and outside the critical region of the Fourier plane where fringe detection is needed to identify binaries ({\it right}).}
    \label{fig:powerspec}
\end{figure}

\begin{figure}[t]
\centering
\includegraphics[height=4cm]{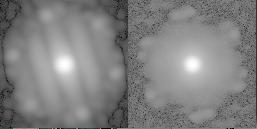}
\includegraphics[height=4cm]{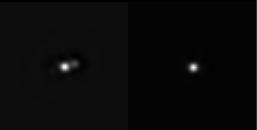}
\caption{\small{Power spectra ({\it left two panels}) and 
reconstructed images ({\it right two panels}) of a 
2:1 flux ratio binary and a 
point source, respectively, modeled (as with the the right
panels of Figs. \ref{fig:speckles} and \ref{fig:powerspec}) using a CFB mask with ffiber core sizes less than half the original speckle sizes.  
Note the lack of a fringe pattern in the power spectrum of the point source
({\it second panel from left}), but very clear fringes for the binary power
spectrum ({\it leftmost panel}) and 
the correlation between fringe orientation
and position angle of the binary ({\it third panel}).
The reconstructed images, essentially inverse Fourier
Transforms of the power spectra, remove the spatial phase differences of the
original speckle pattern and show true images of the sky unadulterated
by the Earth's atmosphere.  This simulation shows 
that the large ring of spikes in the power
spectra caused by CFB fiber core size modulation has no noticeable 
effect on the reconstructed images.}
}\label{fig:recon_image}
\end{figure}

Figure \ref{fig:powerspec} shows the spatial frequency power spectrum derived from stacks of speckle data of a binary star at a separation of twice the diffraction limit, when ({\it left}) using no mask and ({\it center}) using the cohernet fiber bundle mask. The figure illustrates that the effect of the mask (i.e., fiber cellular structure) is confined to a relatively small region in the Fourier $uv$-plane. The dominant fringe pattern seen in each panel shows that the imaged object is clearly a binary star, but the center panel also shows that the effect of the mask is to put spikes of power in a ring whose radius corresponds to the center-to-center distance of the fiber cores on the image plane. Analysis of the detection limits for the masked observation versus the unmasked shows that if one does nothing to counter the effect of the ring of spikes, detection depth is reduced because the background retains the ``waffle'' pattern on the image plane. On the other hand, this problem is mitigated when the 
fiber cores subsample the speckles by at least a factor of 2, 
as we now show
(see Figs. \ref{fig:powerspec} and \ref{fig:recon_image}).

The right panels of Figures \ref{fig:speckles} and \ref{fig:powerspec} 
use the optical DCT data to simulate infrared ($H$ band) observations.
Taking into account the increase in Fried parameter (typical turbulent cell size, $r_0$) from 
0.88 to 1.63$\mu$m (approximately $r_0$ = 43 cm to $r_0$ = 60 cm), 
the speckle images were rescaled to have the correct ratio of speckle size to (long exposure) seeing FWHM (determined by the $r_0$). (In Fig. \ref{fig:speckles}, slightly more than Nyquist pixel sampling of speckles is maintained within a fixed 128$\times$128 pixel format, so the speckle field looks smaller in the zoomed out field-of-view.)
A critical difference in the right image in Figure \ref{fig:speckles} is the reduction of CFB fiber core sizes to 
about half the speckle width. The resulting Fourier domain image (Fig. \ref{fig:powerspec}, {\it right}) shows that the sensitivity to the binary star fringes is maintained but the ring of spikes corresponding to spatial frequencies induced by the CFB core sizes is pushed beyond the diffraction limit boundary, removing its impact on the fringe pattern and restoring the detection limit to the deeper value expected from normal speckle work without fibers. In this domain, reconstructed spatial domain images will have no waffle pattern, as can be seen in 
Figure \ref{fig:recon_image}, made using real speckle data but masked to simulate the CFB with small fiber core sizes.
These simulations provide useful guides for how to scale CFB characteristics to achieve optimal speckle interferometry performance.

\section{PROTOTYPE INSTRUMENT DESIGN}

To test these concepts further, 
we designed a fiber-fed speckle prototype instrument for use at the f/13.5 Cassegrain focus of the University of Virginia's (UVa's) 40-inch telescope at Fan Mountain Observatory (FMO), located near Covesville, VA. We used a 1.7-meter length of Fujikura FIGH-10-500N coherent imaging fiber, which is constructed of 10,000$\pm$1,000 fiber cores, and has an imaging diameter of $460\pm25\mu$m. For the detector we initially chose to use a Raptor Photonics Ninox 
VIS-SWIR 640 InGaAs
infrared camera that was available to us through the Astronomy Department at UVa. In addition to testing fiber-fed speckle imaging, use of the Ninox allowed us to perform speckle imaging in the infrared and test one of the more affordable scientific InGaAs cameras for use in speckle interferometry. However, the Ninox detector is only 640$\times$512 pixels in size, with a pixel size of 15$\mu$m $\times$ 15$\mu$m. Wavelength coverage of the Ninox is $0.4$-$1.7\mu$m. The biggest challenge using this camera is the large amount of noise ---
an RMS of up to 195 e$^-$ in low gain mode, and up to 50 e$^-$ in high gain mode. In addition, the dark current can be as high as 1,500 e$^-$ pixel$^{-1}$ s$^{-1}$ at a detector temperature of $-15^{\circ}$C.

Subsequently, we gained access to an
Andor iXon 888 EMCCD camera. This EMCCD is similar to the ones used in the DSSI [\citenum{horch11}] and NESSI [\citenum{scott16}] instruments. We modified the speckle camera to adapt it to the EMCCD to perform comparison observations. The EMCCD has far better noise performance than the Ninox, with a read noise of $<1$ e$^-$ in electron multiplication (EM) mode, and a dark current of 0.00025 e$^-$ pixel$^{-1}$ s$^{-1}$ at a detector temperature of $-80^{\circ}$C.
Data taken in this configuration allow us a somewhat closer comparison of the fiber results against performance of other established speckle systems (DSSI, NESSI), as well as a direct test of the relative performance of the Ninox camera.

\subsection{Speckle Camera}
The diffraction limit of the FMO 40-inch telescope is $\sim$$0.31^{\prime\prime}$ in $J$-band (1.2$\mu$m) and $\sim$$0.20^{\prime\prime}$ at $\lambda = 0.8\mu$m.  The telescope plate scale is $15^{\prime\prime}$ mm$^{-1}$ and without magnification this gives $0.225^{\prime\prime}$ per pixel on the Ninox. A 30mm focal length collimating lens and a 100mm focal length re-imaging lens were used to give a $\sim$3.3$\times$ magnification onto the Ninox. With single cores from the fiber bundle being $\sim$$4.6 \mu$m, the magnified fiber cores are then $\sim$$15.3\mu$m, or roughly the size of a single pixel. Unfortunately, because the individual fiber cores are not mapped to the individual pixels, a single fiber core will illuminate multiple pixels, leading to some blurring. While our future work will focus on determining the optimal scaling of (speckle diameter):(fiber core):(detector pixel size), 
the present confguration was a compromise given available components at hand.

\begin{figure} [t]
   \begin{center}
   \begin{tabular}{c} 
   \includegraphics[height=3cm]{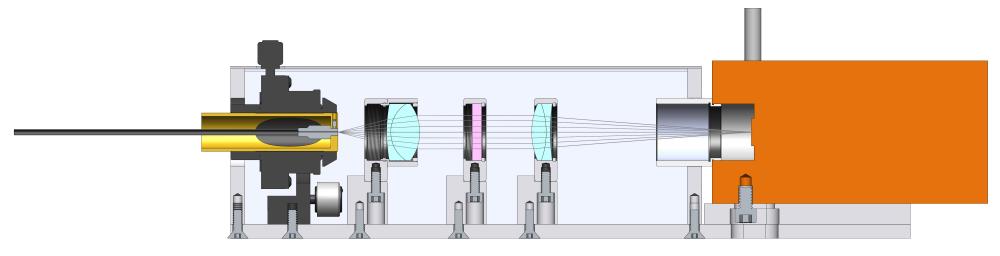}
	\end{tabular}
	\end{center}
   \caption[example] 
   { \label{fig:fiber-fed-speckle-camera-Ninox-cross-section} 
A cross-sectional view of the fiber-fed speckle camera prototype with Ninox camera. The coherent fiber bundle is installed on a Siskiyou fiber translator. Lenses are shown in light blue and the optional $J$-band filter is shown in pink. Light rays are modeled in light gray, and the the Ninox detector is shown in orange.}
   \end{figure}

\begin{figure} [t]
   \begin{center}
   \begin{tabular}{c} 
   \includegraphics[height=5cm]{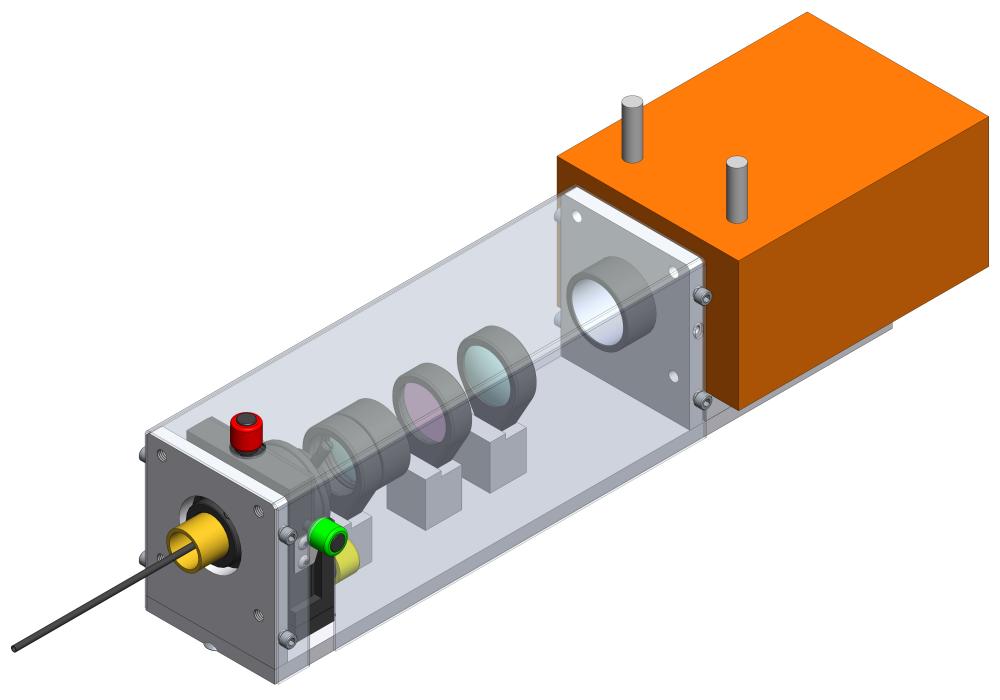}
	\end{tabular}
	\end{center}
   \caption[example] 
   { \label{fig:fiber-fed-speckle-camera-Ninox-isometric} 
An isometric view of the fiber-fed speckle camera prototype with the Ninox detector. 
The cover on the optics box is shown semi-transparent.
}
   \end{figure}

Figure \ref{fig:fiber-fed-speckle-camera-Ninox-cross-section} shows a cross-sectional view,
created in SolidWorks, of the speckle camera as configured for the Ninox detector.
The coherent fiber bundle is attached to a Siskiyou model CTF-5 fiber translator. Light emerging from the end of the fiber is collimated with the 30mm focal length lens.
A filter holder sits in the collimated beam and was either left open or deploying a $J$-band filter. The collimated beam was then focused using a 100mm re-imaging lens. The magnified image of the fiber bundle end is focused onto the Ninox detector, shown in orange in the figure.
A view of the speckle camera with the Ninox is shown in Figure \ref{fig:fiber-fed-speckle-camera-Ninox-isometric}, with the cover shown as semi-transparent, for clarity.
   
We reconfigured the speckle camera to be used with the Andor EMCCD. For comparison to the Ninox results we chose to select optics that would provide roughly the same one-to-one configuration of fiber core size to detector pixel size. We used a 40mm collimating lens and a 100mm re-imaging lens to provide a $2.5\times$ magnification. This produced magnified fiber cores of $\sim9.2\mu$m, which is less than the $13\mu$m pixels of the EMCCD, however this is the closest to a one-to-one relationship we could achieve with available components. For all observations taken with the Andor EMCCD, a filter with a central wavelength of 800nm and FWHM of 40nm was used

\subsection{Fiber Acquisition Unit}
The imaging fiber bundle was connected to the telescope with a small fiber acquisition unit we constructed. 
A Watec video camera mounted to the fiber acquisition unit views the front of the imaging fiber bundle along with an aluminum plate into which the fiber is inserted. The fiber is mounted flush to the face of the aluminum plate which has a polished surface that is slightly angled and reflects the sky to be used for target acquisition and focus.
Guiding during exposure sequences is easily 
effected by watching the readout of the speckle science detector.

\section{OBSERVATIONS}

Observations on the FMO 40-inch telescope with the Ninox camera began in September 2017.  Early test runs with the system made clear that particular care is needed to set the Ninox gain properly to ensure sufficient signal without saturating the detector.  For all Ninox observations taken with the CFB, the 
``High Gain'' mode was utilized, and each
observation consisted of 10,000 individual frames
collected. Initial observations taken in September with the speckle camera directly mounted to the telescope (i.e., without the CFB) used the ``Low Gain'' mode and only 1,000 individual frames were taken.

Whenever possible, we observed stellar binary pairs --- useful as tests of the achieved resolution --- followed by single star observations at a similar position in the sky. The single star observations are used in the reduction of the binary pair data by establishing the power spectrum corresponding to a point source for that particular time, telescope position and instrument configuration. So, for example, in November 2017 we observed using the CFB the well known binary star system HIP 104887, followed right after with observations of the nearby point source HR8215. 
Before each observation of 10,000 frames, preliminary sequences were taken to help with adjustments in the Ninox gain settings to find that which produced the most signal without saturating.
Typical single frame exposure times were either 50 or 100 msec.

Our initial observations with the Ninox camera were taken with no filter. The
broad spectral response of the Ninox -- spanning from 0.4 to 1.7$\mu$m -- means that for those observations the speckles were likely to be significantly wavelength-smeared.
Eventually, in an effort to narrow the spectral range and thereby
make the speckles more coherent, we also took
Ninox observations 
with a $J$-band filter placed in the collimated beam. 
Unfortunately, these $J$-band observations required a higher Ninox gain setting, and this led to an increased dominance of the fixed pattern noise that proved difficult to remove with our existing reduction software.  For this reason, we present only unfiltered Ninox data here.

In the Spring of 2018 we were able to make additional observations of some of the same binary/single star pairings using the Andor EMCCD. To match the Ninox observations, we used 10,000 frame sequences of 50 msec exposures. We also employed an electron-multiplying gain of 200-300$\times$. However, because of the low-noise and low dark current of the Andor camera, for these optical observations we were able to use the $\Delta$=0.04$\mu$m wide filter centered at 0.8$\mu$m. 

Table \ref{tab:Observations} summarizes a fraction of the observations we have made thus far with our prototype CFB speckle camera to date --- those relevant to the results that we present below.  Note that the current 
design of the prototype speckle camera does not lock in
a specific orientation of the detector or fiber bundle on the sky, and all alignments are done ``by eye''.
Due to repeated instrument changes over the course of the observations presented here, we expect that a slop in the on-sky orientation by as much as 5 degrees is possible.

   \begin{table}[t]
\caption{Observation summary. The first four entries, taken with the speckle camera directly mounted, were observed in multiple gain settings in ``Low Gain'' mode with gains of 10, 20, 30 and 50. The EMCCD observations with taken in the EM gain mode.} 
\label{tab:Observations}
\begin{center}             
\begin{tabular}{|l|c|c|c|c|c|c|}
\hline
\rule[-1ex]{0pt}{3.5ex} \bf Target & \bf UT Date & \bf Detector & \bf Mode & \bf Gain & \bf Exposure & \bf Comment \\

\hline
\rule[-1ex]{0pt}{3.5ex}  HIP 104858 & 05 Sept 2017 & Ninox  & direct & 10-50 & 50msec & $0.303^{\prime\prime}$ separation binary \\
\hline
\rule[-1ex]{0pt}{3.5ex}  HR 8178 & 05 Sept 2017 & Ninox  & direct & 10-50 & 50msec & point source for HIP 104858 \\
\hline
\rule[-1ex]{0pt}{3.5ex}  HIP 104887 & 05 Sept 2017 & Ninox  & direct & 10-50 & 50msec & $0.98^{\prime\prime}$ separation binary \\
\hline
\rule[-1ex]{0pt}{3.5ex}  HR 8215 & 05 Sept 2017 & Ninox & direct & 10-50 &  50msec & point source for HIP 104887\\
\hline
\rule[-1ex]{0pt}{3.5ex}  HIP 104887 & 01 Nov 2017 & Ninox  & with CFB & 01 & 100msec & $0.98^{\prime\prime}$ separation binary \\
\hline
\rule[-1ex]{0pt}{3.5ex}  HR 8215 & 01 Nov 2017 & Ninox & with CFB & 01 & 100msec & point source for HIP 104887\\
\hline
\rule[-1ex]{0pt}{3.5ex}  HIP 61941 & 12 Apr 2018 & Ninox & with CFB & 01 & 100msec & $2.72^{\prime\prime}$ separation binary  \\
\hline
\rule[-1ex]{0pt}{3.5ex}  HR 4807 & 12 Apr 2018 & Ninox & with CFB & 01 & 100msec & point source for HIP 61941 \\
\hline
\rule[-1ex]{0pt}{3.5ex}  HIP 61941 & 02 May 2018 & EMCCD & with CFB & 200 & 50msec & $2.72^{\prime\prime}$ separation binary  \\
\hline
\rule[-1ex]{0pt}{3.5ex}  HR 4807 & 02 May 2018 & EMCCD & with CFB & 200 & 50msec & point source for HIP 61941 \\
\hline
\rule[-1ex]{0pt}{3.5ex}  HIP 64241 & 02 May 2018 & EMCCD & with CFB & 300 & 50msec & $0.51^{\prime\prime}$ separation binary \\
\hline
\rule[-1ex]{0pt}{3.5ex}  HR 5013 & 02 May 2018 & EMCCD & with CFB & 300 & 50msec & point source for HIP 64241 \\
\hline
\end{tabular}
\end{center}
\end{table}

\section{DATA REDUCTION}
Reduction of the data followed the same basic methodology that was 
developed for the DSSI 
[\citenum{Horch1}] and used in various projects; a recent example is 
found in [\citenum{horch2}]. 
However, there were relevant changes 
that were implemented for the Ninox data.
We briefly review the
reduction process before explaining those changes. The raw data frames, 
stored as a sequence of FITS images,
are bias-subtracted but 
usually not flat-fielded if the detector is reasonably flat,
since the speckle patterns themselves sample many pixels over the 
entire seeing envelope, and effectively average over (modest) flat
field variations. We then form the autocorrelation of each frame. 
These are added to produce a final, summed autocorrelation for the 
observation. The same procedure is then carried out for a point source 
observation; that is, for an unresolved bright star observed close
in time and close on the sky to the science target. By 
Fourier-transforming these autocorrelations, we arrive at the 
spatial-frequency power spectrum for both observations. The 
diffraction-limited modulus of the object's Fourier trnasform 
can be obtained by dividing the power spectrum of the science target 
by that of the point source and taking the square root. 

This must be combined with the object's phase in the Fourier domain to obtain the diffraction-limited Fourier transform of the object. The
phase function is calculated using the method of bispectral 
analysis [\citenum{Loh83}]. This involves calculating the triple 
correlation of speckle frames and Fourier transforming to arrive
at the bispectrum. In practice, we compute the so-called ``near axis''
subplanes, which can be shown to have a phase that is related to
the derivative of the phase of the object. By extracting this 
derivative and integrating, the phase function is obtained. In our 
case, we use the relaxation 
technique of Meng et al.\ 
[\citenum{Meng90}] to perform the integration in an iterative 
way. Once the modulus and phase are assembled into a complex-valued
function in the Fourier domain, it is low-pass filtered with a
two-dimensional Gaussian function and inverse-transformed to arrive
at the reconstrucetd image. The low-pass filter suppresses 
high-frequency noise, particularly above the diffraction limit,
and the width is chosen so that, on the image plane, the 
full width at half maximum (FWHM) of the reconstructed point spread 
function is comparable to that of an Airy disk at the diffraction limit of the telescope.

For the Andor data described in this paper, no change from the 
above reduction scheme was needed. However, the Ninox camera
has significant read noise, and --- especially when imaging through 
the fiber --- large flat field variations over the scale of several 
pixels. These facts necessitated a slightly more sophisticated 
approach to the pre-processing of data frames. We debiased the
frames as usual, but then hot pixels were also set to zero. 
This left individual frames with a noisy background (albeit one
where the average value was near zero), so we also only included
a pixel in the correlation calculations if it had a value above 
a certain threshold value, chosen to be above the standard 
deviation of the background noise by several times. In most
cases, the brightest features in the frame were still retained,
and so the correlation functions were usable for the purpose 
of reconstructing the image. This method probably introduces 
some bias in the the detection of fainter features (i.e., fainter
secondary stars in the case of binary systems), but this was
judged to be an acceptable trade-off for the engineering data
we have.

\section{RESULTS}

\begin{figure} [t]
   \begin{center}
   \begin{tabular}{c} 
   \includegraphics[height=5cm]{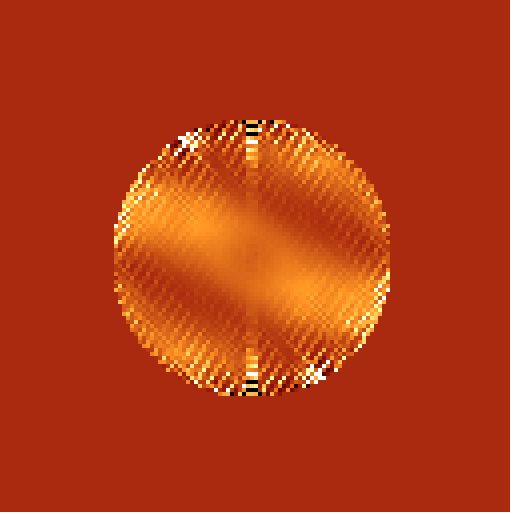}
   \includegraphics[height=5cm]{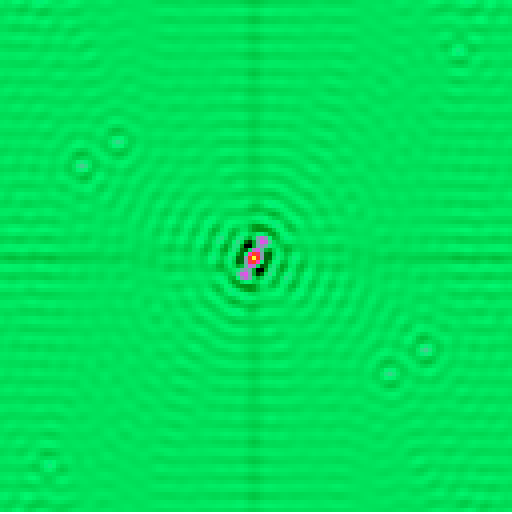}
   \includegraphics[height=5cm]{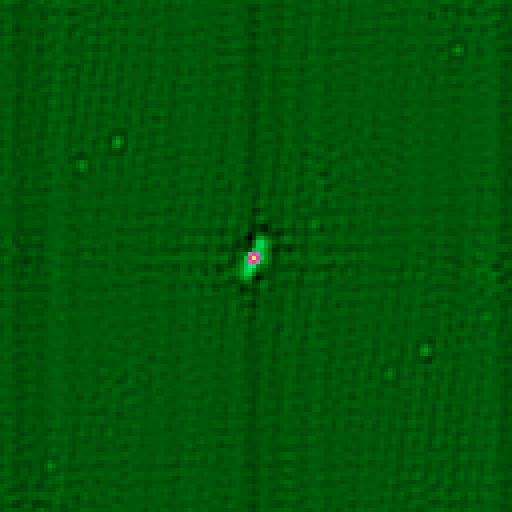}
	\end{tabular}
	\end{center}
   \caption[example] 
   { \label{fig:HIP104858_9-4-2017_ri} 
Speckle imaging results for the binary system HIP 104858 --- a binary with $0.30^{\prime\prime}$ separation --- taken with the Ninox camera directly mounted to the FMO 40-inch (i.e., without a CFB) and unfiltered: (\textit{Left}) Power spectrum from the stack of 4,000 speckle images. (\textit{Middle}) Autocorrelation of the same. (\textit{Right}) Reconstructed image.  All 
panels are North up and East to the right.}
   \end{figure}

To establish a baseline for comparison, Figure \ref{fig:HIP104858_9-4-2017_ri} shows the results of data reduction for frames of HIP 104858 --- a binary with $0.30^{\prime\prime}$ separation --- taken directly onto the Ninox camera, i.e., without a CFB. Note that the separation of this binary star is equivalent to the $J$-band diffraction limit of the 40-inch telescope.  Nevertheless, the power spectrum produced by the 10,000 speckle images obtained clearly shows a fringe pattern, and the reconstructed image and autocorrelation show a companion at a position angle of $\sim$206 degrees with a separation of roughly $0.30^{\prime\prime}$, in good agreement with the expected values of 206.3 degrees and $0.296^{\prime\prime}$ based on our calculations using the orbital parameters from Muterspaugh et al. [\citenum{Muterspaugh08}]. Note that for challenging, small contrast binaries there is often a 180 degree degeneracy obtained in the reconstructed images, and this is the case for this particular target, where a ``double-lobe'' feature is evident in both the autocorrelation space and the reconstructed image space (note that this degeneracy is less severe the other results presented below).
The clear resolution of this $0.30^{\prime\prime}$ 
binary with the 40-inch telescope belies the fact that
these data were taken using the Ninox unfiltered, and
therefore significant signal must have been contributed by 
wavelengths shorter than 1.2$\mu$m, which improved the overall speckle resolution\footnote{While the peak wavelength
of the Ninox response function is approximately 1.2$\mu$m,
the QE curve shows a long tail to $\sim$0.4$\mu$m 
(see http://www.raptorphotonics.com/wp-content/uploads/2016/01/Owl-640.png), while
the transmission of the telescope, which includes
a Baker-Schmidt corrector, likely also selectively curtailed the infrared contribution.}
The primary significance of Figure \ref{fig:HIP104858_9-4-2017_ri} is that it establishes that the Ninox is a viable detector for speckle work.

\begin{figure} [t]
   \begin{center}
   \begin{tabular}{c} 
   \includegraphics[height=4cm]{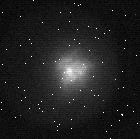}
   \includegraphics[height=4cm]{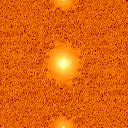}
   \includegraphics[height=4cm]{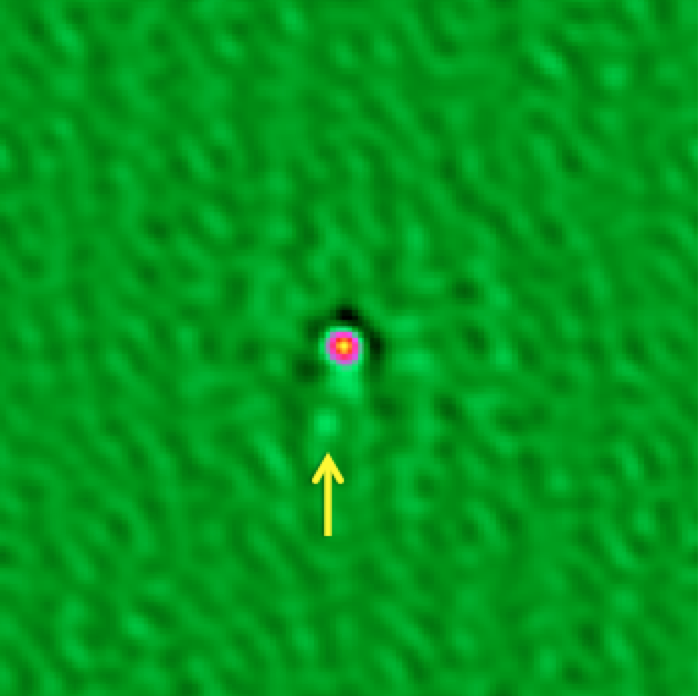}
   \includegraphics[height=4cm]{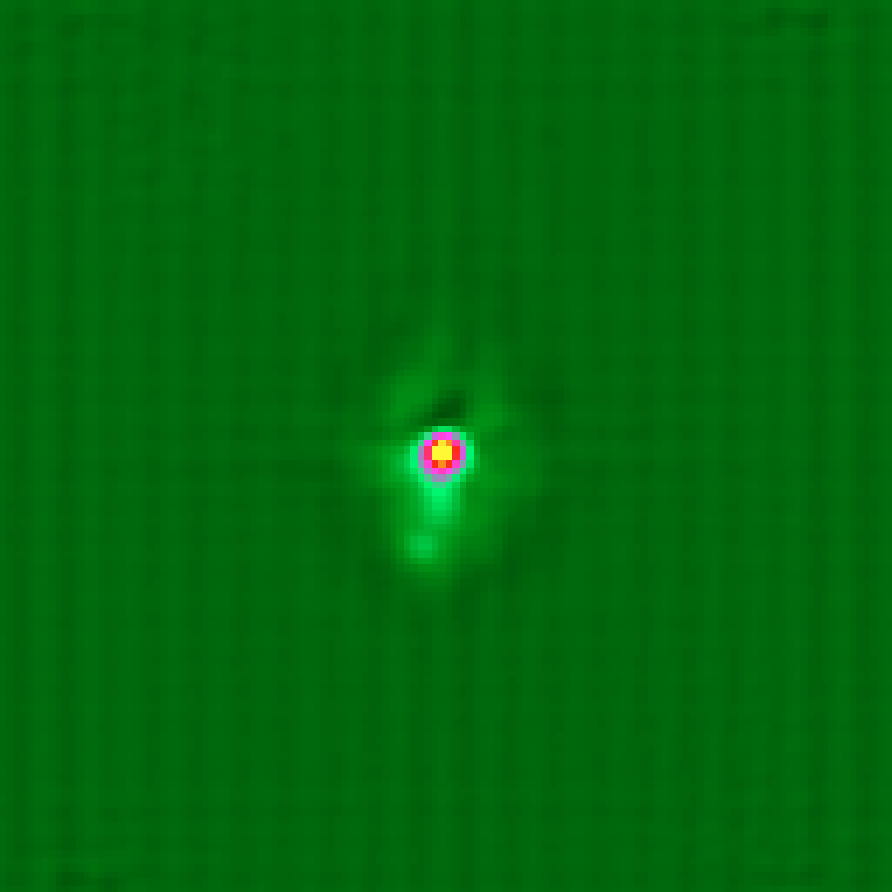}
	\end{tabular}
	\end{center}
   \caption[example] 
   { \label{fig:HIP104887_11-1-2017_ri} 
Observations of the binary system HIP 104887 with the Ninox camera.  ({\it First panel from left}) Sample speckle image taken through the CFB. The many small bright points around the image are pixels with higher dark current that have been overly enhanced by operation at ambient temperature and the high gain setting. In addition, an alternating row striping pattern is present and visible in the raw speckle images taken with the warm detector. ({\it Second panel}) Power spectrum from the stack of 10,000 speckle images.  ({\it Third panel}) Reconstructed image of the binary system taken with the speckle camera through the CFB.  The yellow arrow indicates the location of the secondary, which is at roughly the correct angle and separation.
(Note that cardinal alignment of the speckle system on the sky, with or without the fiber bundle, was not accurately established, and the images are shown with North up and East to the right.).  ({\it Fourth panel}) Reconstructed image of the same binary system taken with the speckle camera 
but {\it without} the CFB, and showing the secondary at the same separation and position angle as in the third panel. }
   \end{figure}

Unfortunately, we were unable to obtain a corresponding
image of HIP 104858 with the CFB in place while this star system was still observable for the season. Instead, the first successful speckle imaging using a CFB was obtained UT 2017 November 01 
of the binary star system HIP 104887.  The secondary
of this binary was expected to have a position angle of 192.5 degrees E of N and a separation of $0.975^{\prime\prime}$ based on our calculations using the orbital parameters from Muterspaugh et al.\ [\citenum{Muterspaugh10}]. Figure \ref{fig:HIP104887_11-1-2017_ri} shows the reconstructed image with the resolved secondary star indicated by the yellow arrow. For comparison, a reconstructed image of the same star system taken with the Ninox camera directly (i.e., without the CFB) --- but a few months earlier --- 
is also shown. Both reconstructed images show a companion at a position approximately consistent with the expected position angle of 192.5 degrees --- though we reiterate that our alignment of the camera system on the sky was not rigorously precise and position angle errors as much as 5 degrees are possible. However, the separation we measure in both images
is in perfect agreement with the expected value.  

Note that the observations with the CFB shown in Figure \ref{fig:HIP104887_11-1-2017_ri} suffered from additional noise
compared to the observations without the CFB due to an error in the control software for the Ninox where the cooling was not activated, and the detector was running at ambient temperature. 
In addition, the long-exposure seeing on the
nights when the CFB observations were made were particularly poor relative to the night when no CFB was used.
Nevertheless, the clear resolution of the subarcsecond binary is obvious in the reconstructed image through the CFB and with nearly the same contrast, despite the fact that the $\lambda = 1.0\mu$m brightness ratio of the two stars in the binary is equivalent to 2.5 magnitudes. 

We also note that the total throughput loss in the system when the CFB was introduced was estimated to be roughly about 1 magnitude. The Fujikura FIGH series bundles nominally have good throughput; including Fresnel losses and no more than 2m fiber lengths for the attenuation, throughputs of 92\% are expected. The higher losses we saw were no doubt in part due to the 
fact that we made no attempt to polish or coat the ends of the CFB after assembly of the prototype system. 
Further work to quantify more precisely the light loss incurred by observations through properly polished CFBs is required.

Overall, these initial results on HIP 104887 proved encouraging. To our knowledge, they were the first demonstration of speckle interferometry through a CFB, yet were obtained despite some challenging circumstances:  (1) the FMO 40-inch telescope has a strong, uncorrected trefoil aberration, likely due to surface imprint from the mirror supports; 
(2) in contrast to the data taken without the CFB, the CFB observations were made through clouds and, with the Ninox detector errantly operating at room temperature, with greatly exacerbated dark current;
(3) the FMO observations were unfiltered, leading to an extremely broad passband (0.4-1.7$\mu$m), which leads to degraded speckle contrast (i.e., $\lambda$/D smearing), (4) the larger than expected throughput loss through 
the CFB, and 
(5) even under proper operating conditions, 
working with NIR detectors adds a strong penalty in
terms of readnoise and additional Poisson noise 
due to dark current.

\begin{figure} [t]
   \begin{center}
   \begin{tabular}{c} 
   \includegraphics[height=5cm]{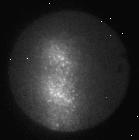}
   \includegraphics[height=5cm]{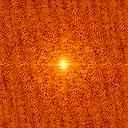}
   \includegraphics[height=5cm]{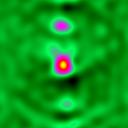}
	\end{tabular}
	\end{center}
   \caption[example] 
   { \label{fig:HIP061941_4-12-2018_ri} 
HIP 61941 observed on April 12, 2018 with the Ninox through the CFB. ({\it Left}) A representative 100 msec raw speckle frame. The outline of the imaging fiber area can be seen. ({\it Middle}) Power spectrum from the stack of 10,000 speckle images ({\it Right})  Reconstructed image showing a companion at $\theta \approx 0^{\circ}$ and $\rho \approx 2.62^{\prime\prime}$.}
   \end{figure}

The last point is made strikingly clear by 
comparisons of observations made with the Ninox versus Andor camera.
Figure \ref{fig:HIP061941_4-12-2018_ri} shows the results of unfiltered Ninox observations of a wider binary pair --- HIP 61941, at $2.72^{\prime\prime}$ spacing. The
example raw speckle image shown in the left panel, where the patches of speckles for each individual star are clearly distinguished, is helpful in illustrating the relative FOV of the CFB on the sky, visible as the lightly exposed area
outlined by the dark black circular frame; as can be seen by comparison against the clearly separated speckles of the binary components, this FOV corresponds to about $7^{\prime\prime}$. 
We measured the secondary companion to lie at a position angle of roughly $0^{\circ}$ and a separation
of $2.62^{\prime\prime}$. Using the orbital parameters of Scardia et al. [\citenum{Scardia07}], we calculate at the time of the observation the position angle was 359.9 degrees East of North with a separation of $2.723^{\prime\prime}$.

\begin{figure} [t]
   \begin{center}
   \begin{tabular}{c} 
   \includegraphics[height=5cm]{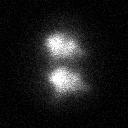}
   \includegraphics[height=5cm]{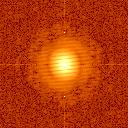}
   \includegraphics[height=5cm]{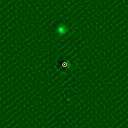}
	\end{tabular}
	\end{center}
   \caption[example] 
   { \label{fig:HIP061941_5-2-2018} 
HIP 61941 observed through the CFB with the Andor iXon EMCCD on May 2, 2018. ({\it Left}) A 50 msec raw speckle frame. ({\it Middle}) Power spectrum from the stack of 10,000 speckle images ({\it Right})  Reconstructed image showing a companion at $\theta \approx 355^{\circ}$ and $\rho \approx 2.73^{\prime\prime}$. (The orientation of these frames is roughly North up and East to the right, to first order.)}
   \end{figure}

As might be expected, the Andor iXon EMCCD, with noise characteristics significantly better than those of the Ninox, 
and imaging a narrower and likely effectively bluer passband,
delivers far better results.
Figure \ref{fig:HIP061941_5-2-2018} shows Andor speckle imaging for HIP 61941, for comparison to the Ninox
imaging of the same system in Figure \ref{fig:HIP061941_4-12-2018_ri}.  Once again, the
two components of the binary are clearly distinguishable by their respective speckle patches in the Andor data, but the quality of the power spectrum and reconstructed image is clearly superior to what is achieved with the Ninox.
The secondary is at a position angle of $\sim 5^{\circ}$ --- within the ``alignment slop'' of the expected position angle of 359.8 degrees at the time of this observation based on the orbital parameters of Scardia et al. [\citenum{Scardia07}]. 
More importantly, we measure a separation of 
$2.73^{\prime\prime}$, consistent with the expected value of $2.728^{\prime\prime}$ from the orbital parameters.

\begin{figure} [t]
   \begin{center}
   \begin{tabular}{c} 
   \includegraphics[height=5cm]{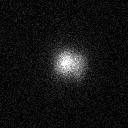}
   \includegraphics[height=5cm]{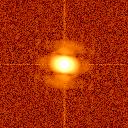}
   \includegraphics[height=5cm]{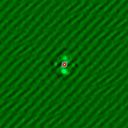}
	\end{tabular}
	\end{center}
   \caption[example] 
   { \label{fig:HIP064241_5-2-2018} 
HIP 64241 observed through the CFB with the Andor iXon EMCCD on May 2, 2018 . ({\it Left}) A 50 msec raw speckle frame. ({\it Middle}) Power spectrum from the stack of 10,000 speckle images ({\it Right})  Reconstructed image showing a companion at $\theta \approx 190^{\circ}$ and $\rho \approx 0.53^{\prime\prime}$. (The orientation of these frames is roughly North up and East to the right, to first order.)}
   \end{figure}

As a more demanding test of the Andor + CFB configuration, we observed the tight binary system
HIP 64241
(Fig. \ref{fig:HIP064241_5-2-2018}).  
As with the Andor observations of HIP 61941 shown
in Figure \ref{fig:HIP061941_5-2-2018}, the long
exposure seeing during these observations was
$1.5$-$2^{\prime\prime}$ (note the large ``speckle field'' sizes in the left
panels of Figs. \ref{fig:HIP061941_5-2-2018} and
\ref{fig:HIP064241_5-2-2018}).
From the orbital parameters by Muterspaugh et al. [\citenum{Muterspaugh15}], we calculate that for the 
night of our observation, a secondary position angle of 192.2 degrees and separation of $0.516^{\prime\prime}$ is expected.
Our observations very clearly resolve the secondary, 
which we measure to lie at 191 degrees (again, 
ignoring alignment errors) with a separation of $0.53^{\prime\prime}$,
well below the seeing limit.
   
A summary of our binary star results compared to the  values expected based on published orbital fitting parameted is shown in Table \ref{tab:ResultsSummary}.  The accuracy of our derived position angles ($\theta$) may be as poor as 5$^{\circ}$ due to our ``by-eye'' alignment of our detectors and fibers on the telescope.

\begin{table}[ht]
\caption{Summary of Results} 
\label{tab:ResultsSummary}
\begin{center}       
\begin{tabular}{|l|c|c|c|c|c|c|c|c|}
\hline
\rule[-1ex]{0pt}{3.5ex}  & & & & \multicolumn{2}{c|}{\bf This Work} & \multicolumn{3}{c|}{\bf Expected}   \\
\hline
\rule[-1ex]{0pt}{3.5ex} \bf Target & \bf UT Date & \bf Detector & \bf Mode & \bf $\theta \ (^{\circ})$ & \bf $\rho \ (^{\prime\prime})$ & \bf $\theta \ (^{\circ})$ & \bf $\rho \ (^{\prime\prime})$ & \bf Ref. \\
\hline
\rule[-1ex]{0pt}{3.5ex}  HIP 104858 & 05 Sept 2017 & Ninox & direct & $\sim$ 206 & 0.30  & 206.3 & 0.296 & [\citenum{Muterspaugh08}] \\
\hline
\rule[-1ex]{0pt}{3.5ex}  HIP 104887 & 05 Sept 2017 & Ninox & direct & $\sim$ 195 & 0.97 & 193.5 & 0.970 & [\citenum{Muterspaugh10}] \\
\hline
\rule[-1ex]{0pt}{3.5ex}  HIP 104887 & 01 Nov 2017 & Ninox & with CFB & $\sim$ 195 & 1.01 & 192.8 & 0.973 & [\citenum{Muterspaugh10}] \\
\hline
\rule[-1ex]{0pt}{3.5ex}  HIP 61941 & 12 Apr 2018 & Ninox & with CFB & $\sim$ 360 & 2.62 & 359.9 & 2.723  & [\citenum{Scardia07}] \\
\hline
\rule[-1ex]{0pt}{3.5ex}  HIP 61941 & 02 May 2018 & EMCCD & with CFB & $\sim$ 360 & 2.73 & 359.8 & 2.728  & [\citenum{Scardia07}] \\
\hline
\rule[-1ex]{0pt}{3.5ex}  HIP 64241 & 02 May 2018 & EMCCD & with CFB & $\sim$ 190 & 0.53 & 192.2 & 0.516  & [\citenum{Muterspaugh15}] \\
\hline
\end{tabular}
\end{center}
\end{table}

\section{SUMMARY AND DISCUSSION}
We have explored the concept of using CFBs in speckle imaging, with an eye toward
the potential for highly multiplexed systems suitable for massive, ground-based, diffraction-limited surveys over large fields-of-view.  Such programs are essential for efficient follow-up of low resolution, space-based photometric transit surveys, not only for identifying false positive or confused detections, but for 
interpreting the range of star/planet system architectures and their dependence on a range of stellar and environmental conditions and for which large statistics are necessary.

Practical considerations for multiplexing speckle imaging suggest that a far more economical alternative to paving telescope focal planes with detectors having pixels small enough to Nyquist sample diffraction-limited speckles (and resigning most pixels at any given time to sampling uninteresting blank sky) is to pipe images of the speckles to a vastly smaller number of detector pixels using CFBs.  
The dimensions of the latest fast readout detectors are large enough to contemplate CFB multiplexing by interesting factors with sufficiently large CFB fields 
of view, and the pace of development will surely soon improve matters for both optical and NIR detectors.

Through simple computer simulations we have demonstrated the impact of the cellular structure of the CFB fiber cores on
speckle imaging.  Not surprisingly, we find that it is important that the fiber core dimensions sufficiently (i.e., Nyquist) sample the speckles. We have yet to fully explore the optimal mapping of fiber core to detector pixel dimensions, but have demonstrated good results even when the fiber cores are of comparable 
diameter to the pixels.

To obtain practical empirical experience and establish the viability of CFB speckle imaging, we have built a prototype speckle imager capable of providing comparison data with and without an intervening CFB.
Successful speckle image reconstructions through a CFB were demonstrated using both an NIR InGaAs-based Ninox camera as well as an optical Andor electron-multiplying CCD, and for binary systems with separations substantially smaller than the native long-exposure seeing and approaching the theoretical telescope diffraction limit.  Position angles and separations for these systems were measured and found to be consistent with expected values (Table \ref{tab:ResultsSummary}).
The superior noise performance of the optical, Andor device delivered the best results, however the opportunity for CFB speckle imaging in the NIR was also demonstrated. 

Our results are even more encouraging when one considers several factors that undoubtedly hindered
the performance of our test system in terms of both throughput and resolution.  Among these factors are: 
The mirror coatings on the FMO 40-inch telescope are severely deteriorated, leading to substantial overall
throughput loss.
The telescope also currently suffers from a severe trefoil aberration most likely caused by failing mirror supports. The coherent fiber bundle we used had unpolished and uncoated surfaces that surely inhibited their net transmission. We also employed a roughly 1:1 matching of the individual fiber core diameters to detector pixel sizes; however, the former are both round and randomly ordered, in contrast to the regular pattern of the (square) detector pixels.  This no doubt resulted in a blurring of the focal plane image. In addition, the observations taken with the Ninox were unfiltered and therefore inordinately pan-chromatic, resulting in a degraded speckle contrast (i.e., $\lambda/D$ smearing). Our very first successful CFB imaging result for the binary system HIP 104887 was made even more challenging by: (1) having been observed through thin clouds, (2) a magnitude difference between the two stars of more than $2$ magnitudes, and (3) the Ninox being operated (errantly) without the cooling system engaged. 

\begin{figure}[t]
\includegraphics[scale=0.80]{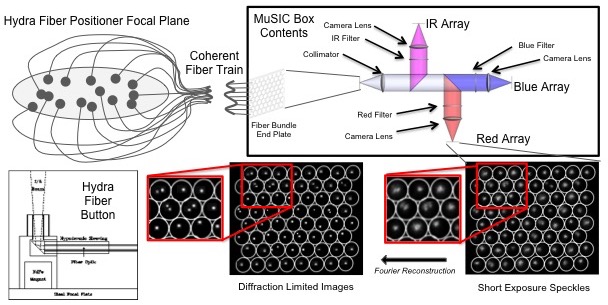}
\caption{
Simplified schematic overview of the MuSIC concept as hosted on the WIYN 3.5-m telescope as an example hosting telescope.  As shown, MuSIC builds on the NESSI design (expending to three wavelength ports) for the camera and exploits the existing WIYN Hydra system for fiber positioning. The inset to the lower left shows a close-up 
view of a Hydra fiber ``button'', which could be adapted to 
mate with a CFB (and miniature fore-optics, if necessary);
the button image is taken from [\citenum{Barden90}].}
\label{fig:MuSIC layout}
\end{figure}

Despite these extensive limitations, we were able successfully to demonstrate that there is no strong physical prohibition to successful speckle imaging through coherent fiber bundles. This is the first proof of concept step on the path toward building a working
multi-object speckle system.  Figure \ref{fig:MuSIC layout} schematically lays out a potential design for such a system.  This Multiplexed Speckle Imaging Camera (MuSIC) is envisioned to couple a network of 64 configurable CFBs in the focal plane of a telescope to a three port (i.e., three wavelength) camera, modeled as an expansion of the successfully-operating NESSI system.  The schematic shows MuSIC working on the WIYN 3.5-m telescope as an example hosting platform, exploiting a modified version of WIYN's existing Hydra fiber positioner, upgraded to fill
its remaining, unused fiber slots with CFBs.  As discussed in \S 2.1, such a system would be capable of simultaneous diffraction-limited imaging of 64 star systems across the 1 degree diameter WIYN FOV, and, through beam splitting, at {\it three} different wavelengths. 
As a potential follow-up device to NASA's TESS mission, 
such a system could cover all of the TESS CTL stars in the Northern Continuous Viewing Zone in about 30 nights of observing --- more than ten times faster than possible with single object systems, even ignoring that the latter would require multiple observing seasons to have enough access to that patch
of sky.

\acknowledgments
The authors would like to thank Garrett Ebelke, Mechanical Engineer, and Charles Lam, Machinist, from the Astronomy Department at UVa for their help in designing and fabricating the speckle camera and fiber acquisition unit used in this work.  We also appreciate useful discussions with Chad Bender and Nick MacDonald.

\bibliography{report} 
\bibliographystyle{spiebib} 

\end{document}